\documentclass[slac_one]{revtex4}
\usepackage{epsfig}
\usepackage{graphicx}
\usepackage{fancyhdr}
\begin{document}
\newcommand{\beq}{\begin{eqnarray}}
\newcommand{\eeq}{\end{eqnarray}}
\newcommand{\non}{\nonumber\\ }

\def \epjc{ Eur. Phys. J. C }
\def \jpg{  J. Phys. G }
\def \npb{  Nucl. Phys. B }
\def \plb{  Phys. Lett. B }
\def \pr{  Phys. Rep. }
\def \prd{  Phys. Rev. D }
\def \prl{  Phys. Rev. Lett.  }
\def \zpc{  Z. Phys. C  }
\def \jhep{ J. High Energy Phys.  }


\title{NLO contributions in the  pQCD approach}
\author{Zhen-Jun Xiao}
\affiliation{The Department of Physics and Institute of Theoretical Physics,
Nanjing Normal University, Nanjing, Jiangsu 210097, P.R. China}

\begin{abstract}
The NLO contributions to some two-body charmless hadronic B meson decays, such as
those from the QCD vertex corrections, the
quark loops and the chromo-magnetic penguins,  have been calculated in the
pQCD factorization approach.
\end{abstract}

\maketitle

\thispagestyle{fancy}

\section{INTRODUCTION}

The two-body charmless hadronic B meson decays play am important role in testing
the standard model (SM) and probing for the new physics beyond the SM.
At present, the key point is how to reduce still large theoretical uncertainty in
evaluating the hadronic matrix elements $<M_2 M_3|O_i|B>$.

During the past decade, many $B \to M_2 M_3$ decays  have been studied by employing the
QCD factorization approach \cite{bbns1}, the soft collinear effective theory (SCET)
\cite{scet1} and the perturbative QCD (pQCD) factorization approach \cite{pqcd1,pqcd2}.
In the pQCD approach, some next-to-leading order (NLO)
corrections, such as those from vertex QCD
corrections, the quark-loops and the chromo-magnetic penguins,  have been included
recently \cite{nlo05,nlo08a,nlo08b},
but much more NLO contributions should be calculated in order to give
more reliable pQCD predictions.

\section{OUTLINE OF THE PQCD FACTORIZATION APPROACH}

In the pQCD approach, the decay amplitude ${\cal A}(B \to M_2 M_3)$ can be written
conceptually as the convolution,
\beq
{\cal A} \sim \int\!\! d^4k_1 d^4k_2 d^4k_3\ \mathrm{Tr}
\left [ C(t) \Phi_B(k_1) \Phi_{M_2}(k_2) \Phi_{M_3}(k_3)
H(k_1,k_2,k_3, t) \right ],
\label{eq:con1}
\eeq
where  $k_i$'s are momenta of light quarks included in each meson, and $\mathrm{Tr}$
denotes the trace over Dirac and color indices. The Wilson coefficient $C(t)$ is
evaluated at hard scale $t$. The function
$H(k_1,k_2,k_3,t)$ describes the hard dynamics and could be calculated perturbatively.
The wave function $\Phi_{M_i}$ describes hadronization of the
quark and anti-quark in the meson $M_i$.
Using the light-cone coordinates the $B$ meson momentum $P_B$ and the two
final state meson's momenta $P_2$ and $P_3$ (for $M_2$ and $M_3$ respectively)
can be written as
\beq
P_B = \frac{M_B}{\sqrt{2}} (1,1,{\bf 0}_{\rm T}), \quad
P_2 = \frac{M_B}{\sqrt{2}}(1-r_3^2,r^2_2,{\bf 0}_{\rm T}), \quad
P_3 = \frac{M_B}{\sqrt{2}} (r_3^2,1-r^2_2,{\bf 0}_{\rm T}),
\eeq
where $r_i=m_i/M_B$. $m_2$ and $m_3$ are the mass of the two final state mesons.
Putting the anti-quark momenta in $B$, $M_2$ and $M_3$ meson as $k_1$, $k_2$, and $k_3$,
respectively, we can choose
\beq
k_1 = (x_1 P_1^+,0,{\bf k}_{\rm 1T}), \quad
k_2 = (x_2 P_2^+,0,{\bf k}_{\rm 2T}), \quad
k_3 = (0, x_3 P_3^-,{\bf k}_{\rm 3T}).
\eeq
Then, the integration over $k_1^-$, $k_2^-$, and $k_3^+$ in
eq.(\ref{eq:con1}) will lead to
\beq
{\cal A}&\sim
&\int\!\! d x_1 d x_2 d x_3 b_1 d b_1 b_2 d b_2 b_3 d b_3 \non &&
\cdot \mathrm{Tr} \left [ C(t) \Phi_B(x_1,b_1) \Phi_{M_2}(x_2,b_2)
\Phi_{M_3}(x_3, b_3) H(x_i, b_i, t) S_t(x_i)\, e^{-S(t)} \right ],
\quad \label{eq:a2}
\eeq
where $b_i$ is the conjugate space coordinate of $k_{iT}$, $S_t(x_i)$ is the
threshold function, and $e^{-S(t)}$ is the Sudakov form factor which suppresses the soft
dynamics effectively.

\section{LO and NLO CONTRIBUTIONS }\label{sec:lo-1}

\begin{figure}[tb]
\vspace{-5cm}
\centerline{\epsfxsize=18cm \epsffile{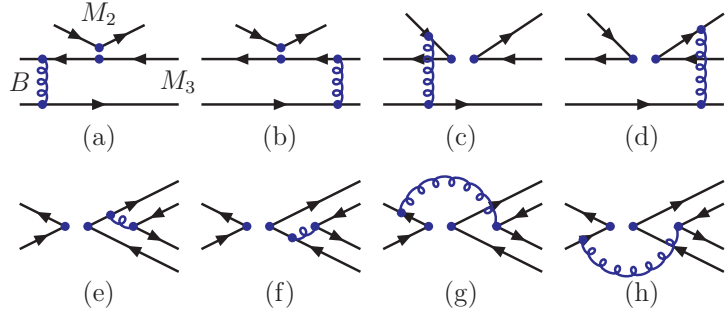} }
\vspace{-16.5cm}
\caption{The typical Feynman diagrams which may contribute to the $B\to M_2 M_3$
decays at leading order.}
\label{fig:pqcd-lo}
\end{figure}

\begin{figure}[tb]
\vspace{-5cm}
\centerline{\epsfxsize=18cm \epsffile{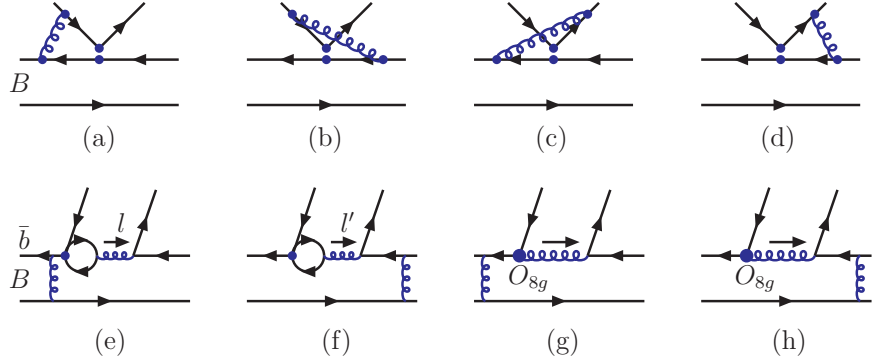} }
\vspace{-15.8cm}
\caption{The NLO contributions in pQCD approach from the vertex QCD corrections (a-d),
the quark-loops (e-f) and the chromo-magnetic penguins (g-h).}
\label{fig:vcqm}
\end{figure}

\begin{figure}[tb]
\vspace{-5.5cm}
\centerline{\epsfxsize=18 cm \epsffile{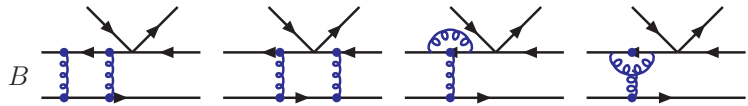}}
\vspace{-19cm}
\caption{The four typical Feynman diagrams, which contributes to the form factors at
NLO level.}
\label{fig:fig3}
\end{figure}

At the leading order in pQCD approach, there are three type
diagrams contributing to the $B \to M_2 M_3 $ decays, the factorizable
emission diagrams, the hard-spectator diagrams and the annihilation diagrams,
as illustrated in Fig.\ref{fig:pqcd-lo}. From the factorizable
emission diagrams Fig.1a-1d, the corresponding form factors can be extracted
by perturbative calculation. For a given decay mode, one can find firstly the individual
decay amplitude by evaluating analytically the corresponding Feynman diagram, obtain
the total decay amplitude by summing up the different pieces and finally calculate
the branching ratios and the CP violating asymmetries.

\begin{figure}[tb]
\vspace{-5cm}
\centerline{\epsfxsize=18 cm \epsffile{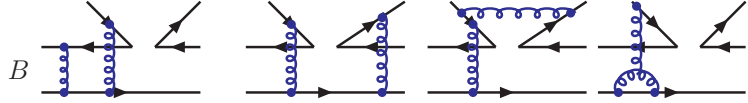}}
\vspace{-19cm}
\caption{The four typical hard-spectator Feynman diagrams, which contributes
at NLO level.}
\label{fig:fig4}
\end{figure}

\begin{figure}[tb]
\vspace{-5.5cm}
\centerline{\epsfxsize=18 cm \epsffile{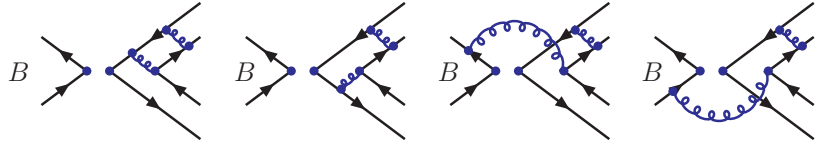}}
\vspace{-18cm}
\caption{The four typical annihilation Feynman diagrams, which contributes
at NLO level.}
\label{fig:fig5}
\end{figure}

When compared with the previous LO calculations in pQCD, the following NLO contributions
should be considered:
\begin{enumerate}
\item
The NLO Wilson coefficients $C_i(M_W)$, and the NLO  RG evolution
matrix $U(t,m,\alpha)$ as defined in Ref.~\cite{buras96}, will be used.

\item
The strong coupling constant $\alpha_s(t)$ at two-loop level will be used.

\item
Besides the LO hard kernel $H^{(0)}(\alpha_s)$, the NLO hard kernel
$H^{(1)}(\alpha_s^2)$ should be included.
All the Feynman diagrams,  which lead to the decay amplitudes proportional to
$\alpha^2_s(t)$, should be considered.
Such Feynman diagrams can be grouped into the following classes:

\begin{itemize}
\item[]{I:}
The NLO contributions from the vertex QCD corrections, the quark-loops and chromo-magnetic
penguins, as illustrated in Fig.~\ref{fig:vcqm}. For chromo-magnetic
penguins,  there are totally nine relevant Feynman diagrams,
but the first two ( Fig.2g-2h ) provide the dominant NLO contributions \cite{o8g2003}.

\item[]{II:}
The NLO contributions to the LO Feynman diagrams (1a,1b) will affect the
extraction of the from factors, as illustrated  in Fig.~\ref{fig:fig3}.
There are totally 13 relevant Feynman diagrams, we here show four of them only.

\item[]{III: }
The NLO contributions to the LO hard-spectator Feynman diagrams (1c,1d), as illustrated
in Fig.~\ref{fig:fig4}. There are totally 56 relevant Feynman diagrams,
we here show four of them only.

\item[]{IV: }
The NLO contributions to the annihilation Feynman diagrams (1e,1h), as illustrated
in Fig.~\ref{fig:fig5}. we here show only four of them.

\end{itemize}
\end{enumerate}

At present, the calculations for the vertex corrections, the quark-loops and
chromo-magnetic penguins have been available \cite{nlo05,nlo08a,nlo08b}.
For more details about the pQCD predictions for the branching ratios and CP violating asymmetries
of $B \to K \eta^{(\prime)}, KK^*, \rho(\omega,\phi)\eta^{(\prime)}$ decays, one can see
Refs.~\cite{nlo08a,nlo08b}.
It is an urgent task to calculate the Feynman diagrams as shown in Figs.~3-5, in order to
provide pQCD predictions at full NLO order.

\begin{acknowledgments}
Work supported  by the National Natural Science
Foundation of China under Grant No.10575052 and 10735080.

\end{acknowledgments}


\end{document}